\documentclass{article}
\usepackage{graphicx,ae}
\usepackage[T1]{fontenc}
\usepackage[latin1]{inputenc}
\usepackage{url}
\usepackage[]{hyperref}
\title{FRACTAL EVOLUTION OF NORMALIZED FEEDBACK SYSTEMS ON A LATTICE}
\author{\\Siegfried Fussy\\
Gerhard Gr\"ossing\\
[.5cm]
{\small\em Austrian Institute for Nonlinear Studies,}\\
{\small\em Parkgasse 9, A-1030 Vienna, Austria}\\
{\small \url{http://web.telekabel.at/ains}}}

\newcommand{\de}{\delta}
\newcommand{\ep}{\epsilon}

\newcommand{\gapp}{\stackrel{\textstyle{>}}{\sim}}
\newcommand{\bea}{\begin{eqnarray}}
\newcommand{\eea}{\end{eqnarray}}
\newcommand{\beq}{\begin{equation}}
\newcommand{\eeq}{\end{equation}}
\newcommand{\ba}{\begin{array}}
\newcommand{\ea}{\end{array}}
\newcommand{\bd}{\begin{displaymath}} 
\newcommand{\ed}{\end{displaymath}}
\begin{document}
\date{}
\maketitle
\vspace{.5cm}
\begin{abstract}
Highly nonlinear behaviour of a system of discrete sites on a lattice
is observed when a specific feedback loop is introduced
into models employing quantum cellular automata \cite{1gr1} 
or their real-valued analogues.

It is shown that the combination of two operations, i.e. i)
enhancement of a site's value when fulfilling a feedback condition
and ii) normalization of the system after each time step, produces
relatively short-lived
spatiotemporal patterns whose mean lifetime can be considered 
as emergent order parameter of the system.
This mean lifetime 
obeys a scaling law involving a control parameter which tunes
the ``fault tolerance'' of
the feedback condition. Thus, within 
appropriate ranges of the systems variables,
the dynamical properties  can be characterized
by a ``fractal evolution dimension'' 
(as opposed to a ``fractal dimension''). 

\end{abstract}

\section{Introduction}

The study of spatiotemporal patterns in dynamical systems has brought
some general insights into the mechanisms of pattern generation and evolution.
Applications exist not only for physical or chemical systems 
exhibiting chaotic behaviour \cite{schuster1}
but also for models of biological networks \cite{schuster2}.

Above all, it has turned out that the systems descriptions via
discrete maps help to avoid complications arising from the analysis of
integro-differential equations \cite{mccauley,oppo}.
One important tool for model building is the use of coupled-map
lattices, either with global couplings as in \cite{kan1}, i.e.
\beq \label{1:map}
x(t+1,j) = (1-\zeta)\,f(x(t,j))+\frac{\zeta}{N}\sum_{i=1}^N f(x(t,i))\,,
\qquad j=1,\ldots,N \,,
\eeq
where $t$ denotes a discrete time step, $j$ the index of elements
of a one-dimensional array and $\zeta$ the coupling constant,
or with local couplings permitting only nearest-neighbour
interaction \cite{kan2}, i.e.
\beq \label{1:cml}
x(t+1,j) = (1-\zeta)\,f(x(t,j))+\frac{\zeta}{2}\{f(x(t,j-1))+f(x(t,j+1))\}
\qquad j=1,\ldots,N \,.
\eeq
The mapping function $f(x)$ is usually chosen as nonlinear function, e.g.
$f(x)=1-a x^2$. Thus the systems dynamics is governed by
a diffusing part (tuned by $\zeta$) and a nonlinear 
transformation (tuned by $a$).

Another approach consists of the simulation of the systems' evolutions by
cellular automata \cite{1wolf}, where time, space and the values of the
state variables (sites) are discrete. As opposed to the coupled-map
models, the source of complexity is the large number
of degrees of freedom and not a nonlinear map.

In this paper we choose a different approach to complex pattern formation: 
we first consider the quantum mechanical analogue of Eq. (\ref{1:cml}), but
with a linear mapping function. 
For the sake of probability conservation, a normalization procedure 
after each time step has to be introduced, which on one hand represents
a source of nonlinearity. On the other hand, it implies a kind of 
nonlocal information spreading all over the array, similar to the
global coupling in (\ref{1:map}). This model was introduced
as quantum cellular automaton (QCA) some years ago and it was also applied
as discretized version of the Schr\"odinger equation \cite{grzeno}.

In its most elementary form the evolution rule reads as
\beq \label{1:origrule}
\Psi(t,j)=\frac{1}{\sqrt{{\cal N} (t)}} 
\{\Psi(t-1,j) + i\de\Psi(t-1,j-1)+i\de^*\Psi(t-1,j+1)\}\,, \quad 
\Psi,\de \in {\cal C}
\eeq
with $\de$ as coupling constant and ${\cal N}$ 
as normalization factor yielding
\beq \label{1:rho} \begin{array}{c}
\rho (t,j) := |\Psi(t,j)|^2 \,,\\
\sum_{j=1}^N \rho(t,j) = 1 \qquad \makebox{for all} \,\, t\,.
\end{array} \eeq
The normalized values $\rho(t,j)$ can be interpreted as probability 
densities for the chosen one-dimensional arrays and are used furtheron
for presentation of the results.
This system 
has been studied thoroughly \cite{nonlocal}.
It exhibits the emergence of exactly predictable coherent
large-scale patterns.
In fact, the effect of the nonlinearity introduced
by normalization is not strong enough to prevent an exact long-term
prediction of its evolution. 

Therefore, to increase the complexity,
we provide the system with a memory in which the sites' values
of the last few hundred time steps are stored. If a specific feedback
condition applied to each site is fulfilled for any site, the value of that
particular site is 
enhanced. Then the array has to be normalized again. Thus the
necessary conservation of probability is guaranteed by confining
all possible (continuous) values of the sites 
within the domain $[0,1]$. Consequently, the arising patterns built
by a group of nonvanishing sites in the spatiotemporal 
plane can be compared with each other,
both when $|\de|<<1$ and $|\de|>>1$.

In the following, we shall focus on only one specific temporal
feedback operation \cite{higher}. 
It will be shown that the complexity of the systems'
behaviours increases considerably due to this second source of nonlinearity.
We think that the arising features of self-organization can serve as 
canditates to understand, at least in principle, some aspects of
neural activity, particulary in those domains where quantum theory
would be necesssary to model neural interactions \cite{grcog}. 

\section{Fractal evolution}

For the purpose of illustration, but without loss of generality, we choose
the parameters of our model as follows.
The systems variables for the ``undisturbed'' 
one-dimensional QCA (cf.
Eq. (\ref{1:origrule})) are the size $N$ (i.e. the number of sites),
the initial values of the state variables $\Psi(0,j)$, and
the mixing parameter $\de=|\de|e^{i\phi},\, \phi \in {\cal R}$.
They are chosen as
\beq \label{2:init} \begin{array}{c}
N=120 \\
\Psi(0,40)=0.1\,,\quad \Psi(0,60)=0.9\,, \quad \Psi(0,100)=0.3 \\
\de \equiv \de_c\cdot (1+i)=0.02 (1+i)\,.
\end{array}
\eeq

The single feedback operation acts on the time coordinate, such that 
a kind of memory is introduced into the system. 
This operation becomes effective after $t_{mem}$ time steps.
Accordingly, the memory of the system consists of an array of the size
$t_{mem}\,\times \,N$ where the sites' values are stored and 
cyclically shifted after each time step 
to keep the most recent values of the array. 
After having reached $t_{mem}$, 
at each time step $t > t_{mem}$ it is checked 
whether the normalized absolute value of each site of the array is similar
to its value at the time $t-t_{mem}\,$.
The similarity check is done by asking whether
the present value lies within a specific range of the past value.
A ``relative interval width'' 
$\ep$ is introduced as crucial control parameter of the model:
\beq \label{2:loop}
\left|\left|\frac{\Psi(t,j)}{\Psi(t-t_{mem},j)}\right| - 1\right|<\ep
\quad \makebox{for all}\,\,j,\, \makebox{and}\,\,t>t_{mem}
\eeq
where $\ep$ varies between $0.0\quad (= 0 \%)$ 
and $1.0\quad (=100 \%)$.

If the condition (\ref{2:loop}) is fulfilled, the value of the site under
consideration is amplified by setting it to $A_{amp}>>1$
\beq
\Psi(t,j)_{enhanced} = A_{amp}\,.
\eeq
Thereafter, the usual normalization procedure is applied.

Note that this kind of feedback operation simulates a selection
process enhancing the values and, consequently, the further evolutions
of those sites which fulfill the similarity condition. We expect that
because of the simplicity of the basic model 
and the additional feedback operation some features of simple neuronal
systems could be modelled, at least with respect to their principal dynamical
structures (cf. \cite{grcog}).

The parameters characterizing the feedback operation are chosen as
\beq \label{2:init_loop} \begin{array}{c}
t_{mem} = 200 \\
\ep=0.041 \, (=4.1\%) \\
A_{amp}=100.
\end{array} \eeq

To have some background against which the arising patterns
can be easily discerned a threshold value $L_{low}$ 
for the probability densities $\rho(t,j)$ is introduced. It both
makes it possible to represent the patterns (consisting of sites with 
$\rho(t,j) \geq L_{low}$)
in appropriate plots or diagrams and to define a physical
lower limit for the operational properties of the system which will be
discussed in detail below. We choose
\beq \label{2:limit}
L_{low}=10^{-6}\,.
\eeq

\begin{figure}
 \centering
\includegraphics*[scale=.6]{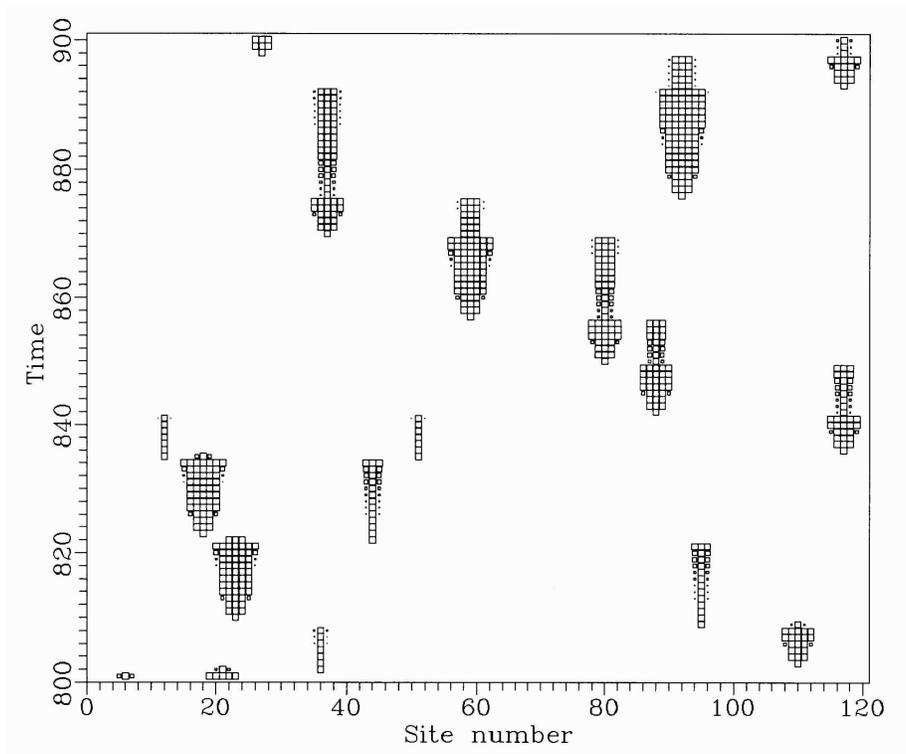} 
\caption{Density plot of a QCA with temporal feedback loop
 accentuating the range of the box values from 
 $L_{low}=10^{-6}$ to $10^{-5}$. Values larger
 than $10^{-5}$ are displayed with the maximal box size. The relative 
 interval width $\ep$ was chosen as $\ep = 4.1\%\,$.}
  \label{fig:1}
\end{figure}
 In Fig. (\ref{fig:1}), the evolution of $\rho(t,j) \geq L_{low}$ 
from time step $t=800$ to $t=900$
with the parameter sets of (\ref{2:init}),
(\ref{2:init_loop}), and (\ref{2:limit}) is displayed as
density plot. The range of values of $\rho$ between $L_{low} = 10^{-6}$ and
$10^{-5}$ is indicated with different box sizes, whereas all values
larger than $10^{-5}$ are displayed with the maximal box size.
Due to the temporal feedback operation 
the continuous evolution of the QCA as studied in Ref. \cite{nonlocal} gets
disturbed after $t_{mem}$. The arising fragments,
operationally defined by a group of spatially and temporally connected
sites whose values are larger than the chosen limit $L_{low}$,
are rapidly distributed all over
the plane, irrespective of the patterns' locations due to 
the initial configuration. This feature can be understood in the
following way: if a site fulfills the feedback condition,
its value is strongly amplified relative to the neighbour's values.
Thus it represents a nucleus for further evolution. After
a certain time, another site will fulfill the time loop condition
and will be enhanced, whereas the previous evolution will be damped after
normalization. This behaviour continues until the damping process
due to the appearance of new nuclei leads to an extinction of the
evolving pattern which now remains as a fragment in the density plot.

\begin{figure}
\centering
\includegraphics*[scale=.6]{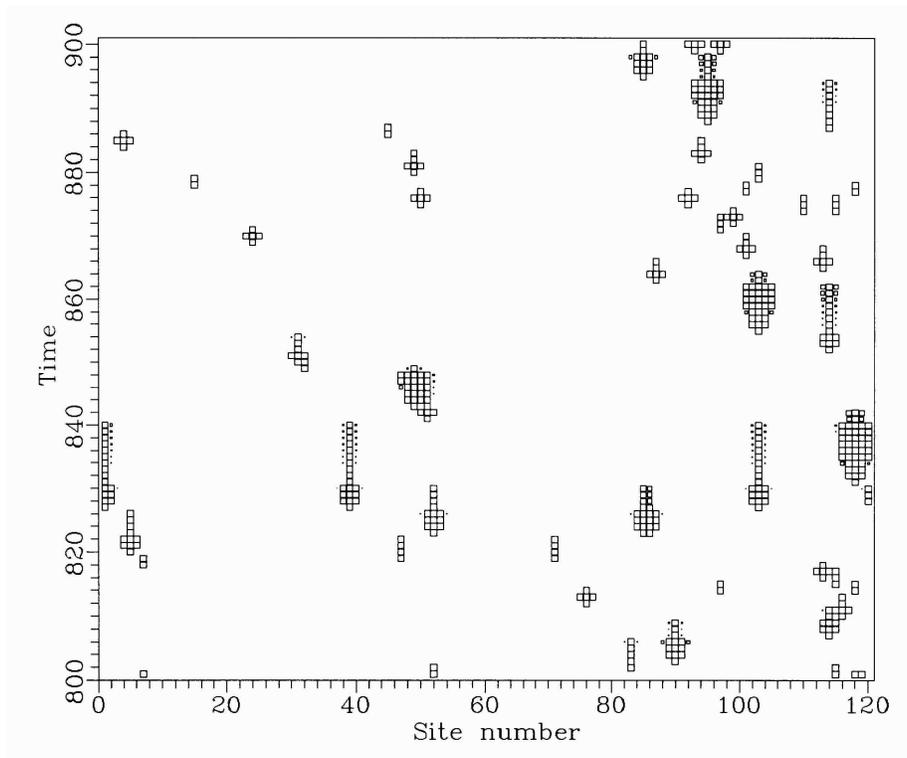} 
 \caption{Density plot with the same conditions as in Fig. (\ref{fig:1}), but with
$\ep=25\%\,$. The fragmentation of the spatiotemporal patterns
clearly has increased.}
  \label{fig:2}
\end{figure}
This behaviour indicates a certain
correlation between $\ep$ as measure for the probability of
generating new nuclei on one hand,
and the lifetimes $\tau$ of the fragments which depend
on the frequency of the appearance of new nuclei on the other. This
correlation can be seen from the comparison of Fig. (\ref{fig:1}) with
Fig. (\ref{fig:2}), where $\ep$ has been enlarged up to
$\ep = 25\%\,$.
The fragmentation of the evolving patterns has 
clearly increased, and the mean lifetime of a single
pattern has decreased. Therefore, a characterization of
this new statistical property of the system 
by taking the mean lifetime of the patterns
$\tilde{\tau}=\tilde{\tau}(\ep)$ is justified. 
Now the important role of a threshold
value for $\rho$ becomes clear: operationally, $\tilde{\tau}$, being
a kind
of ``order parameter'' of the system, has to be constrained 
by physical limits.

By varying $\ep$ between $0.12\%$ and $90\%$,
the plots were analyzed with regard to the mean lifetimes $\tilde{\tau}$ of
the patterns. A ``pattern recognition'' program was 
written to count and analyze each bounded
fragment up to $t=5\times 10^4$ and to calculate $\tilde{\tau}$ (simply as 
arithmetic mean) and its mean error $\sigma$. The latter includes also
a systematical error, mainly due to ambiguities in choosing
well defined fragments. With the help of parallel
visual checks of the plots it was estimated as roughly $5\%$ of $\tilde{\tau}$.

\begin{figure}
\centering
\includegraphics*[scale=.6]{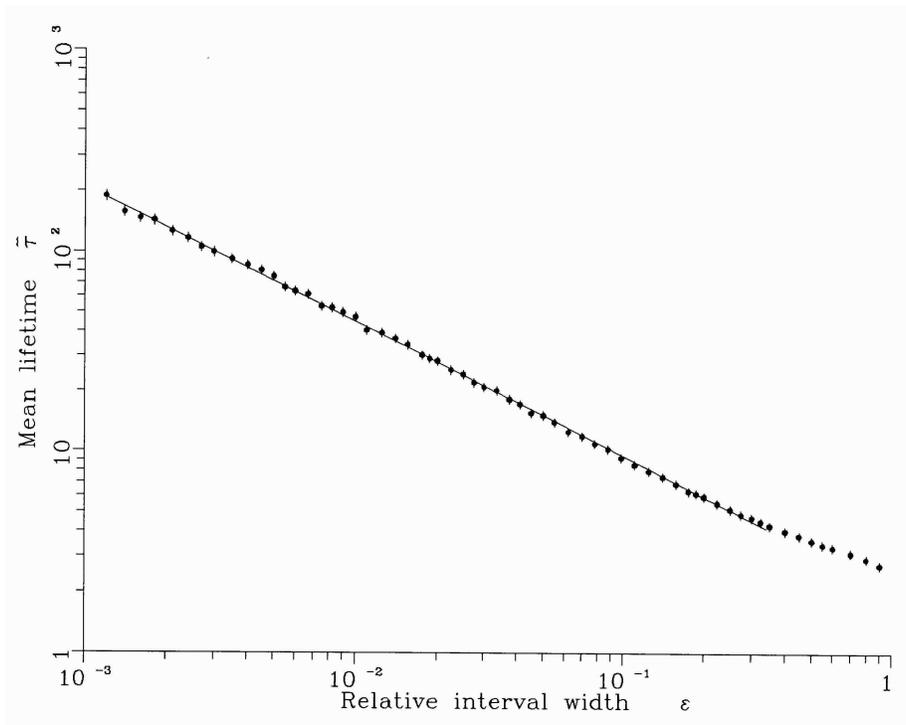} 
 \caption{Log-log plot of the mean lifetimes $\tilde{\tau}$ 
of the fragments versus the
relative interval width $\ep$. Systems parameters
are taken from Eqs. 
(\ref{2:init}),(\ref{2:init_loop}), and (\ref{2:limit}). 
The data are fitted with a power-law
function. Thus a ``fractal evolution dimension'' characterizing
the pattern formation can be derived.}
  \label{fig:3}
\end{figure}
The result can be seen in Fig. (\ref{fig:3}), where the plotted data are shown
in a log-log diagram. As a surprising result, the data produce, at
least up to $\ep \approx 35\%$, a linear
function, indicating a kind of effective ``fractal dimension'' of 
the patterns independent
of the relative interval width $\ep$. The fit was performed with
the power-law function 
\beq \label{2:fit}
\tilde{\tau}=a\cdot\ep^{b}\quad
\eeq
where
\beq \label{2:result}
a=2.00 \pm 0.03,\quad b=-0.675 \pm 0.005
\eeq
with a $\chi^{2}$ of $14.9$ for $52$ degrees of freedom.

We want to emphasize the difference in our use of $b$ as opposed to the 
usual measure of a fractal dimension of an object. Usually, the
fractal dimension exhibits an ``object's'' 
scale invariant features, whereby
the process of its determination does not change the intrinsic
properties of the ``object''.
In our case, the scale invariance of the fragmentation of the evolving
patterns is generated by the repeated 
application of different values of the relative 
interval width. We call this dynamical 
fragmentation process a ``fractal evolution''
of the system. The scale invariant property of the emerging order parameter
of the system, characterized by the mean lifetimes of the patterns, 
will be denoted as ``fractal evolution dimension''
($D_E$). In analogy to the usual algorithm \cite{3man,3tak}, one 
gets the ``fractal evolution
dimension'' $D_E$ as 
\beq
D_E=1-b=1.675\quad \pm 0.005\quad .
\eeq

Note also that the fragmentation ceases (i.e. the lifetime of the
pattern at the beginning goes to infinity) as $\ep$ goes to zero
(i.e. when the precision of the comparison reaches a maximum). 
This behaviour is plausible, because the probability for the feedback
condition
getting fulfilled is very small for small values of $\ep$, so that
the evolution of the QCA remains nearly undisturbed.

\begin{figure}
\centering
\includegraphics*[scale=.6]{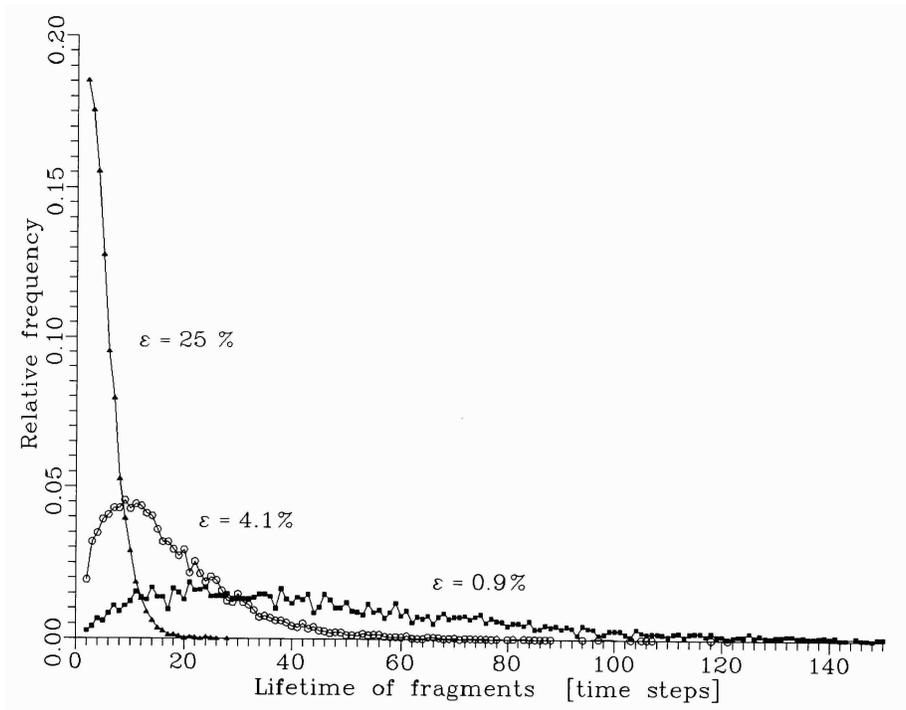} 
 \caption{Distribution function of the patterns' lifetimes obtained for 
$t_{mem}=200$ and for
$\ep=0.9\%\,,\ep=4.1\%\,,$ and $\ep=25\%\,$, respectively. Characteristic
asymmetric functions are observed.}
  \label{fig:4}
\end{figure}

The deviation of the data points in Fig. (\ref{fig:3}) from the linear
function for $\ep \gapp 35\%$
can be understood at least qualitatively by looking at the shapes
of the relative frequencies for the different lifetimes $\tau$ belonging
to a specific $\ep$. In Fig. (\ref{fig:4}),
they are displayed for three different values of $\ep$, i.e for
$\ep = 0.9\%\,, 4.1\%\,,$ and $25\%\,,$ respectively. 
Note that with the default parameter values as introduced above
no fragments of length $\tau=1$ can occur: assuming the 
enhancement of some site
at location $j$ and time $t$, the value of $\rho(t,j)$ will be about 1.0
after normalization, whereas the other sites will be damped by a factor
of about ${\cal N} \approx A_{amp}^2 = 10^{4}$. 
Even if, at the next time step, 10 different
sites would fulfill the feedback condition and thus become enhanced, the
normalization factor would be of the 
order ${\cal N} \approx 10\times 10^{4}$.
Therefore, $\rho(t+1,j)$ would indeed be reduced to about $10^{-5}$, but 
it would still remain 
above the chosen threshold of $L_{low}=10^{-6}$. Thus, a
minimum length of 2 for each arising fragment is guaranteed.

The distributions are all asymmetric, showing
poisson-like ``tails'' proportional to $e^{-\tau}$, but for large
$\ep$ the cut-off at the lifetimes' value of $\tau=2$ becomes effective. 
Thus, no fragments with $\tau < 2$ can contribute, and
the mean value of those lifetimes will be shifted to a higher value.

We also report briefly on the 
variation of the systems parameters \cite{nextp}.
It turned out that fractal evolution is practically invariant under $i)$
variation of the phase of the coupling
parameter $\de$ (cf. Eq. (\ref{2:init})), $ii)$ the initial values of
the state variables, and $iii)$ the choice of other time intervals 
(than from $t_{mem}$ to $t=5\times 10^4$) for the pattern analysis.

The variation of the remaining variables in 
(\ref{2:init}), (\ref{2:init_loop}), and (\ref{2:limit}) in most cases 
also yields
fractal evolution, but with shifted values of $a$ and $b$ in (\ref{2:result}).
Deviations from fractal evolution are observed in cases of ``transient
states'' of the system according to specific values of the parameters.
In all cases where such a deviation
occurs also the shape of the distribution function for $\tilde{\tau}$
deviates from that described above. 

Finally, even the rule (\ref{1:origrule})
itself can be changed in an appropriate way without loosing the property
of fractal evolution. 
For example, one can model a transition to
a ``semi-classical'' system by restriction to real-valued parameters
and couplings, i.e. 
\beq \label{1:realrule}
\Psi(t,j)=\frac{1}{\sqrt{{\cal N} (t)}} 
\{\Psi(t-1,j) + \de[\Psi(t-1,j-1)+\Psi(t-1,j+1)]\} 
\eeq
with $\Psi,\,\de \in {\cal R}$.
Also in this case a fractal evolution can be
observed. This is opposed to the effects of the restriction
 to real values in the undisturbed case \cite{nonlocal},
where a ``real-valued QCA'' always leads to trivial final states. 

\section{Conclusions}

An implementation of a specific temporal feedback operation in
normalized dynamical systems on a lattice 
with a linear mapping function provides
new and surprisingly rich evolutionary properties.
Specifically, the temporal behaviour
of QCA evolution with feedback loops selecting and amplifying particular
regularities was investigated. 
We have found that these regularities are 
independent of the initial point configurations, of the phase
of the coupling parameter, and of the choice of the time
interval whithin which pattern analysis is performed.

Most importantly, we have found a strict scaling behaviour of the emerging
spatiotemporal patterns: on a log-log scale, lifetimes $\tilde{\tau}$ of the
patterns' fragments are a strictly linear function of the relative
interval width $\ep$ fed back into the loop comparing the amplitudes
at different evolution times. In other words, there exists a ``fractal
evolution 
dimension'' $D_E$ of the patterns whose value is independent of the particular
choice of $\ep$. As $\ep$ is the {\underline {relative}} 
interval width, $D_E$ is a quantity characterizing all the emerging patterns
irrespective of the absolute values of the intervals in the temporal 
feedback loop. 
Above all, the
observed features are not restricted to one specific model. It
can be concluded that discrete,
normalized feedback systems with linear mapping
functions generally exhibit the property of fractal evolution \cite{ott}.

Apart from being a ``toy model'' to study pattern generation in
nonlinear systems, the observed features of our model
are of particular interest for possible applications of QCA
like modelling processes in the brain as discussed in 
Ref. \cite{grcog}. As much of the encoding and decoding of neural signalling
is a function of the ``lifetimes'' of neural firing, we have seen that
our model is of interest because the system generating the characteristical 
``temporal patterns'' does not have to operate with a ``knife's edge''
precision. On the contrary, our model gives an example of the unimportance
of precision in some pattern recognition tasks: It just takes any two
values for the average lifetimes $\tilde{\tau}$ at different relative interval
widths $\ep$ to determine the ``fractal evolution dimension'' $D_E$
typical for 
the feedback loop in the QCA, and thus to obtain a measure
of $\tilde{\tau}$ under arbitrarily high precision $\ep$ (that is, within
physical limits, of course). Such features may be relevant for real-time
analog signal processing (as opposed to digital ``all-or-nothing''
processing with correspondingly high precision requirements) which allows
for high effective noise levels and fault tolerances of neural 
computations \cite{4bia}.



\begin{thebibliography}{99}
\bibitem{higher}
QCA with additional feedback loops can be denoted as ``higher order QCA''
in analogy to perturbation theory as opposed to an ``undisturbed'', 
``zero order QCA''.
\bibitem{nextp}
Details will be discussed in a forthcoming paper.
\end{thebibliography}
\end{document}